%% file: ms.tex
\documentclass[12pt,preprint]{aastex}

\input brmacro.tex
\input psfig.sty

\slugcomment{}

\shorttitle{Abell 1068}
\shortauthors{McNamara, Wise, \& Murray}

\begin{document}

\title{The Insignificance of Global Reheating
in the Abell 1068 Cluster: Multiwavelength Analysis}

\author{B. R. McNamara\altaffilmark{1}}
\affil{Dept. of Physics \& Astronomy, Ohio University, Athens, OH 45701}
\email{mcnamarb@ohio.edu}

\author{M. W. Wise}
\affil{Massachusetts Institute of Technology/Center for Space Research}
\email{wise@space.mit.edu}

\author{S. S. Murray}
\affil{Harvard-Smithsonian Center for Astrophysics}
\email{ssm@head-cfa.harvard.edu}

\altaffiltext{1}{Visiting Astronomer, Kitt Peak National Observatory.
NOAO is operated by AURA, Inc.\ under contract to the National Science
Foundation.}

\begin{abstract}
In this paper and in a companion paper, Wise, McNamara, \& Murray (2003), hereafter referred
to as WMM, we present a detailed, multiwavelength
study of the Abell 1068 galaxy cluster, and we use this data
to test cooling and energy feedback models of galaxy clusters. 
Near ultraviolet and infrared
images of the cluster show that the cD galaxy is experiencing
star formation at a rate of $\sim 20-70\msunyr$ over the past $\lae 100$ Myr.
The dusty starburst is concentrated toward the nucleus of the cD galaxy 
and in filamentary structures projecting $60$ kpc 
into its halo.  The {\it Chandra} X-ray image presented in WMM reveals 
a steep temperature gradient that drops from roughly 4.8 keV beyond 120 kpc
to roughly 2.3 keV in the inner 10 kpc of the galaxy where the
starburst peaks.  Over 95\% of the ultraviolet and H$\alpha$
photons associated with the starburst are emerging from regions
cloaked in keV gas with very short cooling times ($\sim 100$ Myr), 
as would be expected from star formation fueled by 
cooling condensations in the intracluster medium.  
However, the {\it Chandra} spectrum
is consistent with but does not require cooling at a rate of $114-145\msunyr$, 
factors of several below the rates found with the {\it ROSAT} observatory.
The local cooling rate in the vicinity of the 
central starburst is $\lae 40 \msunyr$, which is consistent with the
star formation rate determined with $U$-band and infrared data.
We consider energy feedback into the intracluster medium by
the radio source, heat conduction, and 
supernova explosions associated with the starburst.
We find that energy feedback from both the radio source and
thermal conduction are inconsequential in Abell 1068.
Although supernova explosions associated with the starburst
may be able to retard cooling in the inner 10 kpc or so of the cluster
by $\sim 18\%$ or so, they are incapable of maintaining the cooling
gas at keV temperatures.
Finally, we present circumstantial evidence for the contrary view that
at least some and perhaps all of the star formation may have been
fueled by an interaction between the cD and one or more
companion galaxies.

\end{abstract}


\section{Introduction}

Galaxy clusters frequently possess central cusps of bright  
X-ray emission from relatively high density, low temperature
gas.  The radiative cooling time of this gas approaches 
several hundred million years in the centers of some clusters, which is
much shorter than their ages.   
Unless the thermal energy losses are replenished by a robust heating agent, 
the gas will cool to low temperatures and sink to the center of the cluster.  
The natural, long-term repositories for the cooling
gas are atomic and molecular clouds and  stars.
Over the past 20 years, a large body of observational
evidence has emerged in support of the existence of such a repository,
although at substantially lower levels than expected. 
The centrally dominant galaxies (CDGs) that lie at the centers
of so-called ``cooling flows'' often harbor cool interstellar media traced by 
nebular line emission (Heckman et al. 1989, Voit \& Donahue 1997), 
neutral hydrogen (O'Dea, Baum, \& Gallimore 1994, Taylor 1996), 
molecular gas (Edge 2001, Edge et al. 2002.
Jaffe \& Bremer 1997, Jaffe, Bremer, \& Van der Werf 2001,
Donahue et al. 2000, Falcke et al. 1998), 
and star formation (Crawford et al. 1999, Cardiel, Gorgas, \& Aragon-Salamanca
1998, McNamara \& O'Connell 1989).
In addition, a strong correlation exists between
the occurrence and strength of star formation 
in CDGs and the X-ray cooling rates measured with the {\it Einstein}
and {\it Rosat} observatories (see McNamara 1997, 2002 for reviews).  
The star formation rates are, however, typically only several to several
tens of solar masses per year compared to cooling rates of
several hundred solar masses per year.  Furthermore, 
the molecular gas reservoirs that often exceed
$\sim 10^{10}\msun$ in some clusters (Edge 2001) 
would in most cases fill to capacity in only several tens of Myr,
which is too short a timescale to account for the cooling mass in
a long-lived  cooling flow.  Our inability to account
for this apparent mass continuity violation is the crux of the
so-called ``cooling flow'' problem.

Observations of galaxy clusters obtained with the {\it Chandra} and
{XMM-Newton} observatories have dramatically changed our
view of cooling flows.  High resolution images obtained
with {\it Chandra} have revealed remarkably complex gaseous structures 
in the vicinity of the central galaxies (McNamara et al. 2000, 2001,
Fabian et al. 2000, and many others).  In addition,
both the moderate to high resolution spectra obtained with {\it Chandra}'s
ACIS camera and the XMM-{\it Newton}'s Reflection Grating Spectrometer
do not show the signatures of steady cooling below 2 keV 
throughout the central regions of clusters 
(McNamara et al. 2000; Molendi \& Pizzolato 2000; David et al. 2001, Fabian et al. 2001, Peterson et al. 2001, B\"ohringer et al. 2002). 
Apart perhaps from the very inner
regions surrounding the complex X-ray structures,
the gas can be adequately modeled as a single temperature plasma.
Thus, standard inhomogeneous
cooling flow models with gas cooling to low temperatures
throughout the central $\sim 100$ kpc of clusters at a steady rate are
apparently incorrect.
Nevertheless, the cooling upper limits from {\it Chandra} and XMM-{\it Newton} 
do not rule-out cooling in the inner regions of clusters 
at rates ranging between a few and a few hundred solar masses per year
(Peterson et al. 2003).  These limits are often intriguingly close to 
the observed levels of
star formation in the central cD galaxies.  Furthermore, the 
sites of star formation occur in the  vicinity of complex X-ray structures
where the cooling time of the gas is less that $3\times 10^8$ years 
(McNamara et al. 2000, Blanton et al. 2002, McNamara 2002).  
The local cooling rates surrounding the star formation regions
are within factors of several of the star formation rates, as would
be expected were star formation  fueled by cooling.

The outstanding issue now is whether one or more heating agents 
capable of preventing wholesale cooling over cosmological timescales
can be identified.  One such heating agent is the radio source.  
Most cooling flows contain an audible radio source in their central galaxies
(Burns 1990), and they frequently appear to be interacting 
with the ambient keV gas.  For example, large cavities or bubbles
have been found in the X-ray emission of many clusters
that  were evidently created by strong interactions between
the central radio sources and the surrounding gas (B\"ohringer et al. 1993,
Carilli et al. 1994,  McNamara et al.
2000, Fabian et al. 2000).  The energy
liberated in $PV$ work and by bulk lifting may be sufficient to
retard or quench cooling in some objects,  although
the details of these processes are not understood
(Heinz, Reynolds, \& Begelman 1998, Reynolds, Heinz, \& Begelman
2002, Basson \& Alexander 2002, Nulsen et al. 2002, David et al. 2001, 
Fabian et al. 2002, Kaiser \& Binney 2003, 
Br\"uggen et al. 2002, Br\"uggen \& Kaiser
2002, Br\"uggen 2003, Soker et al. 2001, 
Brighenti \& Mathews 2002, Churazov et al. 2001, Churazov
et al. 2002, Quilis et al. 2001, De Young 2003).  An additional source
of heat may be thermal conduction from the hot outer layers
of clusters (Rosner \& Tucker 1989,
Fabian, Voigt, \& Morris 2002, Narayan \& Medvedev 2001,
Voigt et al. 2002, Soker, Blanton, \& Sarazin 2003, Zakamska \& Narayan 2003). 
Acting in concert, a cycle of radio-induced outflow and heat 
inflow via thermal conduction 
(Ruszkowski \& Begelman 2002) may regulate cooling at the
levels of cold gas and star formation observed in CDGs 
(McNamara et al. 2000, Edge 2001,
Edge et al. 2002). $Chandra$'s sharp images now permit the cooling 
rates to be measured
on the same spatial scales as star formation, providing
a direct test of cooling and heating models.  Applying this test and exploring
the consequences of feedback in the Abell 1068 CDG is
the purpose of this paper.

We present deep $U$, $R$, and H$\alpha$ images
of the central region of the $z=0.1386$ cluster Abell 1068, and
we compare them to new $Chandra$ imaging discussed in
detail in WMM.  Abell 1068 was selected among several of the most distant
clusters with X-ray evidence for strong cooling flows from the
$ROSAT$ Brightest Cluster Survey (Allen et al. 1992).
The observations were intended to study the star formation
strength and morphology of distant CDGs using the
stellar population-sensitive $U$-band and the H$\alpha$ feature.  
Throughout this paper, we assume ${\rm H_0}=70~{\rm km ~s^{-1}~Mpc^{-1}}$,
$\Omega_{\rm M}=0.3$, $\Omega_\Lambda =0.7$, $z=0.1386$,
an angular diameter distance of 505 Mpc, and that 1 arcsec = 2.45 kpc.

\section{Optical Observations}

The optical observations were obtained with the Kitt Peak
National Observatory 4m telescope's T2KB CCD camera at prime
focus on 1 February, 1995.  This configuration delivered
a plate scale of $0.47$ arcsec/pix.  
Images were taken through the standard $U$ plus liquid
Copper Sulfate red-leak blocking filter, Gunn $R$, and
two intermediate bandwidth
filters centered on the redshifted $H\alpha+$[N II] features and
their adjacent continuum. Total exposure times were 3600 seconds, 
800 seconds, 1400 seconds, and 1200 seconds, respectively.
The target images were taken in ``short scan''
mode, which shifts charge in the CCD during an
exposure in order to improve the flat field quality of the images.
The target images were then individually flat-fielded using
twilight sky images, the bias level was subtracted
from each image, and the images were combined into the
science images used in our analysis.

\section{Data Analysis}

\subsection{Structure of the Central Galaxy}

The central 1.2 arcmin ($176 \times 176$ kpc) of the cluster is shown in the
$U$ and $R_{\rm g}$ bands in Fig. 1.  The light in the
center of the cluster is dominated by the CDG
and two dozen or so fainter galaxies.  Both the $U$-band and $R$-band 
images show a wisp of light projected between the CDG's nucleus
and a bright galaxy  15 arcsec to the south-west.
A fainter, distorted galaxy is projected
33 arcsec to the north-west of the CDG's nucleus.  In addition,
there are several faint knots of light along a line 
between this galaxy and the CDG.  These features, in addition
to the CDG's blue nucleus, will be discussed here in detail.

In Fig. 2 we present the surface brightness profiles of the
CDG in $U$ and $R_{\rm g}$.  The profiles were
constructed using elliptical annulae. 
The annulae have ellipticities of $\simeq 0.4$ and position angles of 
$\simeq 130\deg$ matching those of the $R$-band isophotes at radii 
between $\sim 10-15$ arcsec.
The surrounding galaxies were masked from the images
in order to isolate the CDG light prior to the analysis.
The surface brightness profile in each band was flux calibrated using
Landolt standards. The calibrated magnitudes were then transformed
from the laboratory frame to the rest frame using K corrections
from Coleman, Wu, \& Weedman (1980).  The K corrections in
$U$ and $R_{\rm g}$ are $0.51$ mag and $0.13$ mag respectively.
The Galactic foreground extinction toward Abell 1068 is very small,
thus no correction for foreground extinction was applied.  The $U$ and $R_{\rm g}$
profiles include statistical error bars (imperceptible in all but
the outer $U$-band points),  and systematic error confidence intervals
(dashed lines) on the U-band profile.  The method for determining
the confidence intervals is discussed in McNamara \& O'Connell (1992).

The $R_{\rm g}$ profile follows the
$R^{1/4}$ law  between 1.5 arcsec and 33 arcsec
($3.7-81$ kpc) and over 5 magnitudes of surface brightness. The
surface brightness profile rises abruptly 
a half magnitude above the $R^{1/4}$ profile
(straight line in Fig. 2) at a radius of 33 arcsec (81 kpc)
and surface brightness of 25 mag
per square arcsecond and remains there to the limiting
radius of our photometry.  This departure above the
$R^{1/4}$ profile is the signature of an 
envelope, which is the defining characteristic of a cD galaxy (Schombert
1986).  Therefore, we will use the terms cD and CDG interchangeably 
when describing Abell 1068's central galaxy.  The U-band profile likewise follows an $R^{1/4}$
profile between 4 arcsec and 33 arcsec (10 kpc and 81 kpc). 
Beyond this radius the surface brightness profile likewise
rises above the $R^{1/4}$ law.  
Apart from the blue core, these properties are 
typical of central cluster galaxies within a redshift $z\lae 0.1$ 
(Porter et al. 1991).  

Within a radius of 4 arcsec (10 kpc) the $U$-band profile brightens by
nearly a magnitude compared to the $R^{1/4}$ profile.
In Fig. 3 we show the $(U-R)_{\rm K,0}$
color profile derived from the surface brightness profiles in Fig. 1.  
To establish a point of reference, the normal rest-frame colors for a cD galaxy
generally lie within the range of $2.3-2.6$.  The nuclear colors
tend toward the red end of this range (Peletier et al. 1990).  The cD's 
$U-R$ color becomes approximately 0.9 magnitudes bluer than normal in
the central 4 arcsec.  Beyond 4 arcsec the color profile remains 
roughly constant 
at a value of $(U-R)_{\rm K,0}\sim 2.3$ well into the halo of the galaxy.

\subsection{Blue Morphology of the Central Galaxy}

In addition to the the blue nucleus, there are several
anomalously blue structures projected onto the halo of the cD galaxy.
In Fig. 3 we present a $U$-band image of the central galaxy after
subtracting a smooth stellar  model.
The residual image clearly shows the two
dozen or so galaxies and the wispy
features seen in Fig. 1.  We placed letters on the
map to identify features of interest.  The blue  nucleus, 
{\it A}, is at the center of the figure indicated with a ``$+$''.
Eight arcsec to the north-west
at {\it B} lies an arc-like feature with a faint light bridge
to the nucleus.  Extending from {\it B}, a wisp of light, {\it C}, curves
22 arcseconds to the south-west toward the nucleus of
a bright galaxy, {\it D}.  A fainter galaxy, {\it E}, is seen 36 arcsec
to the north-west of the nucleus.  What appear to be either
small galaxies or debris near {\it F} are seen between {\it A} and galaxy 
{\it E}.  

Fig. 3 includes a color map of this region on the
same spatial scale superposed onto the $R_{\rm g}$
isophotal contours.  Relatively normal colors are
shown in white and the unusually blue regions are shown in
blue.  As expected, the region within a 4 arcsec radius
of the nucleus is  blue, as are features
{\it B}, {\it C}, and {\it F}.  Note the normal nuclear
colors of galaxies {\it D} and {\it E}.  Colors, magnitudes, and
average surface brightnesses of these regions are given in
Table 1.  The bluest region outside the nucleus is the bright
arc {\it C}.  Adopting $(U-R)_{\rm K,0}\sim 2.3$ as
the typical color of giant elliptical star light,
region C is 0.2--0.6 magnitudes bluer than normal.
Likewise, features B and F are $\sim 0.2$ magnitudes and
$\sim 0.4-0.5$ magnitudes bluer than normal, respectively.

The integrated magnitudes of galaxies {\it D} and {\it E} (Table 1)
 are $R_{\rm g}=16.94$ and $R_{\rm g}=18.53$, respectively,
corresponding to absolute magnitudes
$M({\rm R_g})=-22.1$ and $M({\rm R_g})=-20.6$,
or equivalently, $1.3L^*$ and $0.3L^*$,
respectively.  They are clearly massive galaxies
whose colors are consistent with those of giant ellipticals
or perhaps spiral bulges.  We see no star formation
within these galaxies, and they appear to be
interacting with the cD.  If they are triggering star formation
in the process, galaxies {\it D} and {\it E} are more
likely to be disrupted spiral bulges rather than giant
ellipticals (cf., \S 7).  

The arc-like features {\it B} and {\it C} are noteworthy in
that we cannot exclude the possibility that one or both are
gravitational arcs, rather than local star formation.
The more compelling gravitational arc would be {\it B}, as it appears
to be centered on the cD galaxy, which is presumably the local
center of mass. However, inspection of Figures 3 and 4 shows
that this feature is associated with a ``hook'' of H$\alpha$
emission (discussed in detail in the next section), 
which would be a surprising coincidence were the
feature indeed a gravitational arc. The more
likely interpretation of this unusual feature is star formation
triggered by an interaction with the radio source shown
in Figure 4.  Feature
{\it C} represents a poorer case for a gravitational
arc because its radius of curvature is much larger than
its radial distance from the cD, and it is not centered on the cD.  
On the other hand, strong H$\alpha$ emission is absent
toward this feature, and toward the apparent debris trail at {\it E}
and {\it F}, which weakens to some degree the case for star formation
in these regions.  However, the absence of strong  H$\alpha$ emission
toward arc {\it C} and the debris trail is probably
a consequence of their low optical 
luminosities and correspondingly low
local star formation rates.  In any event, the true interpretation
of the arc-like features would have only a minor impact on the
conclusions of this paper. 

\subsection{Nebular Emission}

In Fig. 4 we present an H$\alpha+ [{\rm N} II]$ map of the cD.  
A bright, extended emission-line nebula
is centered on the cD and is somewhat more extended
than the blue nucleus.  The nebula
is irregular in shape with a tongue of emission extending to the north-west
terminating in a knot of emission at the blue, arc-like feature {\it B}.
The H$\alpha$ luminosity is $\gae 10^{42}~\ergsec$ (Allen et al. 1992).
The surface brightness profile of the emission nebula is shown 
as crosses in Fig. 2, scaled arbitrarily to show its spatial
extent relative to the $U$-band excess.  It is sharply peaked on the
cD's nucleus but drops below detectability at a radius
of 10 arcsec, where the northern tongue of emission terminates.
In this respect, Abell 1068 is similar to other cooling flow clusters 
harboring luminous emission nebulae  
(eg., Heckman et al. 1989, Crawford et al. 1993).

\subsection{The Radio Sources}

Both the cD and the bright galaxy {\it D}
to the south-west are 20 cm {\it FIRST} survey radio
sources (White et al. 1997, Gubanov \& Reshetnikov 1999). 
For reference, the 5 arcsec resolution radio map is superposed  
on the $U$-band contours in Fig. 4.
Their radio fluxes are $8.71\pm 0.14$ mJy and $10.87 \pm 0.14$ mJy,
respectively. The cD's radio source is marginally 
resolved and extends along position angle $\simeq 150\deg$ in the
direction of the tongue of H$\alpha$ emission to the north-west.
Both the radio source and the tongue of H$\alpha$ emission terminate
at the location of the bright blue arc {\it B}.  
A close spatial relationship between
the radio source, nebular emission, and knots of star formation
is a common pattern in powerful radio galaxies and in cooling
flows.  Abell 1068 seems to fit this pattern.

\subsection{Cold Gas and Dust}

The infrared source F10378+4012 lies at the position of the cD.  
The $IRAS$ Faint Source Catalog lists its 60 and 100 micron fluxes as 
577 mJy and 958 mJy respectively.  Combined, they imply
a total infrared flux of $3\times 10^{-11}~{\rm erg~s^{-1}~cm^{-2}}$
(cf., Wise et al. 1993).   At Abell 1068's distance, the
corresponding total infrared luminosity is
$L_{\rm FIR}(40-500\mu {\rm m})\ge 1.5\times 10^{45}~{\rm erg~s^{-1}}$.  
Assuming the infrared light is being emitted by dust grains, the 
dust temperature and mass are 31 K and
$2\times 10^8\msun$ respectively. The location and
spatial extent of the infrared emission region is unknown, but is
probably associated with the starburst.
Assuming the dust is distributed in a foreground
screen of 10 kpc radius with scattering properties similar to dust in our
galaxy, the visual extinction and color excess
would be $A_{\rm V}\sim 4.3$ and $E(B-V)\sim 1.4$ magnitudes,
respectively.  Extinction with these properties would
be readily apparent in our optical images, and 
the color excess would be much larger
than $E(B-V)=0.39$ found by Crawford
et al. (1999) based on the anomalous Balmer decrement in
the emission nebula.  This inconsistency  implies that
the dust is either confined to very small scales, or is 
broadly distributed throughout the cluster.

Greater than $4\times 10^{10}\msun$ of molecular gas has been detected
at millimeter CO ($2\rightarrow 1$) and CO($1\rightarrow 0$)
features (Edge 2000), and in the 2$\mu$ lines of molecular hydrogen (Edge 
et al. 2002).  The apparently close spatial association of the 
2$\mu$ emission and H$\alpha$ emission in this and other cooling flows
(Donahue et al. 2000) suggests that the molecular material,
and probably the dust are closely associated with the starburst.  
The gas to dust ratio
implied by the $IRAS$ and CO data must be $\gae 200$, which is in line with 
other clusters (Edge 2001).   Edge pointed out that 
Abell 1068's high infrared luminosity and its abundance
of molecular gas are properties in common with those 
of the luminous infrared galaxies (Sanders \& Mirabel 1996).

\subsection{Star Formation in the Central Blue Structure}

Anomalously blue colors are observed in many cooling flow CDGs.
Although most are associated with star formation, 
blue color excesses can be caused by metal poor stellar populations
or emission from active nuclei. The normal central colors of CDGs 
lie in the range $(U-R)_{\rm K,0}\sim 2.3-2.6$ (Peletier et al. 1990).
A normal giant elliptical galaxy
becomes bluer with increasing radius by  by roughly two tenths
of a magnitude per decade in radius (Peletier et al. 1990).
By comparison, normal central cluster galaxies 
have somewhat shallower gradients, and sometimes may redden with increasing radius  
(Mackie 1992, McNamara \& O'Connell
1992).  A gradual blueing with increasing radius is
thought to originate in populations of decreasing average
stellar metallicity, decreasing age, or a decreasing
level of dust absorption.  The general trend for ellipticals to 
possess red, metal enhanced or dusty cores is inconsistent
with the blue core in Abell 1068.  Therefore, it 
cannot be explained straightforwardly as a metallicity effect within
the usual context of giant ellipticals.

Emission from an active nucleus is a possible origin for the central blue
colors, but it would have difficulty explaining
Abell 1068's overall properties.  The nuclear blue emission is distributed
almost symmetrically about the nucleus, and there is no evidence
for anisotropies associated with beamed nuclear radiation.
Assuming this light is being scattered
isotropically from a central source, we would expect to see a 
bright central peak and faint halo structure with a large
contrast ratio due to the inefficient scattering of photons 
by electrons and dust.  This is not seen, nor
is a point source seen in the $Chandra$ X-ray image.  
Finally, we and others (Allen et al. 1992, Allen 1996, and
Crawford et al. 1999) have shown that the blue 
core is associated with nebular emission whose line ratios
are a closer match to H II regions than AGN. 
Taken together, Abell 1068's properties are
consistent with ongoing massive star formation; 
we will proceed with this assumption.

\section{Star Formation History of the cD Galaxy}

\subsection{The Nuclear Starburst}

Our primary objective is to 
explore the well known relationship
between star formation and the centrally-peaked, cluster X-ray emission.
Chandra's high spatial resolution provides for a direct comparison
between the star formation regions and X-ray emission 
on the same arcsec spatial scales.  This
unique capability provides a crucial test of the cooling paradigm,
which we apply in this section of the paper.

In order to estimate the star formation rate, or more precisely,
the luminosity mass of the young stellar population, we 
measured the fraction of the $U$-band light
emerging from the blue population in the inner 4 
arcsec radius ($\simeq 10$ kpc) by modeling the $U$-band 
light of the host galaxy 
using an $R^{1/4}$-law surface brightness profile with a
softened core.  The fraction of light, $f_{\rm AP}$, 
contributed by the blue ``accretion population'' was found by subtracting
the model $U$-band image of host galaxy from the real image,
leaving the accretion population in residual.  
This analysis shows that the accretion population contributes up to
75\% of the $U$-band light in the inner arcsec, and averages 
$12\%$ over the inner  4 arcsec radius region.  
We determined the mass of the accretion population as 
$M_{\rm AP}=M/L(U)_{\rm AP}f_{\rm AP}L(U)$, where
$M/L(U)_{\rm AP}$ is the $U$-band mass-to-light ratio of the accretion
population, and $L(U)$ is
the total $U$-band luminosity of the central blue region.
The stellar population histories
are based on the Bruzual \& Charlot (1993) population models with
the Salpeter initial mass function and Solar abundances. 
We found the $U$-band luminosity of the accretion population alone
to be  $L(U)_{\rm AP}\equiv f_{\rm AP}L(U)=2\times 10^{10}~{\rm L_{\odot}}$, 
before correcting for extinction.

Measuring the intrinsic color of the accretion population
is the most difficult aspect of this analysis.
The color constrains the population's age and history,
which in turn is used to estimate its mass-to-light ratio
by comparison to stellar population models. The accuracy of the measurement
is limited by our ability to separate the light of the accretion
population from the surrounding glare of the background cD galaxy.
While the $U$-band signal from the accretion population is relatively
strong and therefore somewhat easier to model, 
the fraction of the total light contributed by the
accretion population in the $R$-band is only $\sim 5\%$.  Such a
small contribution to the total light implies that a
small error in the galaxy model can lead to a substantial misestimation
of the $R$-band  flux, and a correspondingly large error in the
accretion population's color.  Adding to these difficulties,
the light from the accretion population is
heavily concentrated toward the nucleus where its light fraction
varies dramatically with radius.  
Furthermore, the true central surface brightness
profile of the background population at $U$ is poorly determined.
With these caveats in mind, we find a 
probable range for the central accretion population color of 
$(U-R)_{\rm AP}\sim -0.1$ to $-0.3$.  

This color range is broadly consistent with 
two star formation histories: continuous
star formation with a range of ages, and a simple, instantaneous
burst (cf., Table 2).  For the blue end of the color range,
the instantaneous burst and continuous star formation histories
are indistinguishable, having ages 
of $\sim 6-10$ Myr and similar mass-to-light ratios. The
red end of the color range is consistent with ongoing star formation
for $\sim 100$ Myr, or an aging burst that occurred roughly 40 Myr ago.
The continuous star formation history brackets
a 10 Myr to 100 Myr time span at rates between $40 \msunyr$
and $16\msunyr$, respectively.  The corresponding accretion population
masses would be  $4\times 10^8~{\rm M}_\odot$ and 
$1.6\times 10^9~{\rm M}_\odot$, respectively.   The colors are
equally consistent with a fading, instantaneous burst that
occurred between $6$ Myr and $40$ Myr ago, with  corresponding accretion
population masses of $4\times 10^8~{\rm M}_\odot$ and
$10^9~{\rm M}_\odot$, respectively.   
The uncertainty in our color measurement does not
permit a precise reconstruction of the star formation
history.  However, the additional constraints discussed below
suggest the starburst has consumed a small fraction
of the available molecular fuel, and may be in the
early stages of a more extended episode of star formation.

\subsection{Off Nuclear Star Formation}

An inspection of Fig. 3 shows a series of blue regions
lying at angular distances of between a few and several tens of
arcsec from the nuclear starburst.  These regions are described
in \S 3.2 and their apparent colors and magnitudes are given
in Table 1.  The large arc at location {\it C} is 
two tenths of a magnitude bluer than
the nucleus, while
the smaller arc at {\it B} is two tenths of a magnitude 
redder than the nucleus.  The colors
include light from the background cD galaxy and do not represent
the intrinsic colors of the blue population themselves.  Using
procedures similar to those described above,  the intrinsic colors
of the off nuclear blue features range between $U-R_{\rm K,0}\sim -0.2$
to  $U-R_{\rm K,0}\sim 0.1$.  This color range brackets those
of the nucleus and extends redward.  
The total $U$-band luminosity of the off-nuclear
regions is $5\times 10^9~{\rm L}_\odot$, only one quarter that of
the nuclear $U$-band luminosity (Table 2).

In order to interpret
the colors and luminosities in terms of an average star formation
history, we combined
the off-nuclear blue light from regions {\it B, C,} \& {\it F},
and summarized their properties in Table 2.
The color range 
of the off nuclear population is consistent with 
a short duration burst that occurred
between 7 and 50 Myr ago, with a corresponding luminosity mass
between $\sim 0.9-3\times 10^8\msun$, 
and with constant star formation at a rate of
between $3-5\msunyr$ that has been ongoing 
for $100-500$ Myr.  The physical appearance of the population
and its possible association with an interaction between the
cD and two of its neighbors favors the burst history.

\subsection{Far Infrared Star Formation Rate}

The far infrared luminosity 
in the direction of the cluster was found to be
$L_{\rm FIR}(40-500\mu {\rm m})\ge 1.5\times 10^{45}~{\rm erg~s^{-1}}$
in \S 3.5.  Assuming all of the infrared flux is processed ultraviolet
light from the starburst, the star formation rate based on the
global Schmidt law for spiral and starburst galaxies (Kennicutt 1998)
is $68 \msunyr$, with a formal uncertainty
of $\sim 50\%$.  This figure is only 1.5--3.5 times larger than
our estimates in Table 2, and is well within the uncertainties
of the measurements.  Taking the optical and infrared star formation
rates at face value, the star formation region is largely
obscured, and most of its radiation is emerging in the infrared.

\section{Star Formation Fueled by Cooling}

If star formation is being fueled by cooling condensations 
in the X-ray gas, we expect to find the coolest gas with
the shortest cooling time in close proximity to the 
regions of star formation.
In order to test this condition,
we compare the blue regions with the temperature
map in Fig. 5,
and to the isobaric cooling time profile, Fig. 12b in WMM.  
The temperature
map, shown with the same centering and spatial scale as the optical images
in Figs. 1, 3, \& 4, shows a great deal of structure.  
Three distinct concentrations of cool gas are evident.  
The coolest concentration shown in dark blue is centered on the cD. 
The second extends to the south-east of
the cD, and a third extends to the west.

The coolest gas in the cluster at $\simeq 2.5$ keV 
is projected on the cD's inner $\simeq 6$ 
arcsec coincident with the nebular emission and
the region from which $\sim 75\%$ of the $U$-band light
from the young stellar population is emerging. 
Fig. 12b in WMM shows that the isobaric cooling time is $\sim 10^8$ yr
in the inner 10 kpc, and then rises to $\sim 10^9$ yr at a radius
of 100 kpc, reaching a value of $\sim 10^{10}$ yr at a radius of
$\sim 200$ kpc.  The gas in the central starburst
region has the shortest isobaric cooling time in the cluster 
$t_{\rm c}\simeq 9\times 10^7$ yr.  
This trend is
seen in other objects such as the Hydra A and
Abell 1795 clusters (McNamara et al. 2000, Fabian et al. 2000).

The situation with the off-nuclear star formation is
less clear. Regions {\it B}
and {\it C}, which contain $\sim 23\%$ of the
blue light, are located within the tongue of cool gas extending
to the west of the cD.  The gas surrounding this region
has a short cooling time of $t_{\rm c}\sim 5\times 10^8$ yr.
On the other hand, the blue knots of star formation
at {\it F} are in a higher temperature region with a cooling
time exceeding $\sim 10^9$ yr.  The connection
of this material to cooling is less compelling,
although it accounts for only $2\%$ of the excess $U$-band
luminosity.
This material may instead be associated with debris that has been
stripped from the nearby galaxy at {\it E}. 
It is worth noting that no obvious star formation is  
associated with the tongue
of cool gas that extends $20-30$ arcsec south-east of the cD,
although we cannot exclude the possibility of a faint, diffuse
population of young stars there.

\subsection{Comparison Between Cooling Rates and Star Formation Rates}

In WMM, we reported several emission models
capable of reproducing the X-ray spectra of the
gas in the central 100 kpc of the cluster.  The  maximum
total cooling rate of roughly $1000\msunyr$ is 
comparable to those found in earlier $ROSAT$ analyses 
(eg., Allen 2000). This model is not 
well fit to the data.  Furthermore,
the cooling stops at 2 keV, with very little
cooling flux at lower energies.  This behavior, which is now
commonly seen clusters, indicates that little gas 
is cooling to low temperatures, or much of the gas
is cooling to low temperatures with
a low radiative efficiency (Fabian et al. 2000).

Spectral models exist that fit the X-ray spectra reasonably well
that require gas to cool
to very low temperatures, albeit at much smaller rates (WMM).
These ``full cooling'' models constrain the maximum
level of cooling below X-ray temperatures to be
$114-145 \msunyr$ (depending on the assumed foreground
hydrogen column density) to a radius of $\sim 200$ kpc,
well beyond the star formation radius.
The deprojected full cooling rate in the central
5 arcsec of the cD galaxy, where most of the star formation is
occurring, is $\dot m=45\pm 20\msunyr$.  This local cooling rate agrees with the
the measured range of constant star formation $\dot s\sim 20-68\msunyr$.
The high value is the far infrared star formation rate
discussed earlier. The probable  mass of the accretion population ranges
between $4\times 10^8~{\rm M_\odot}$ and $2\times 10^9~{\rm M_\odot}$,
a small fraction of the molecular gas mass, $4\times 10^{10}
~{\rm M_\odot}$ (Edge 2000).  Considering the star formation analysis
above, this would suggest that star formation will continue over
an extended period of time, and not in a short
duration  burst.  If the molecular gas is 
fueling the star formation at the level of  $15-70 \msunyr$, 
star formation will be quenched in roughly one Gyr, assuming
no new cold material is accreted.  It is
therefore likely that star formation is in its early
stages.

\section{Feedback \& Reduced Cooling in the Clusters's Core}

It became apparent from the earliest cluster observations with
{\it Chandra} (McNamara et al. 2000; Fabian et al. 2000) 
and {\it XMM-Newton} (Molendi \& Pizzolato 2001; Peterson et al. 2001)
that the cooling rates reported from the previous generation of
X-ray observatories were either vastly overestimated, or
that cooling must be occurring with a very low radiative efficiency
(Fabian et al. 2000, B\"ohringer et al. 2002).
The former hypothesis implies the gas is being maintained
at X-ray temperatures by a heating agent.  Possibilities include
the radio sources in cD galaxies (Rosner \& Tucker 1989),  
heat conduction from the hot outer layers of clusters
(Narayan \& Medvedev 2001), and
supernova explosions associated with star formation
in the cooling gas (Silk et al. 1986).
The interest in radio sources
has been revived by {\it Chandra}'s discovery
of X-ray cavities or bubbles associated with radio
sources in clusters (see McNamara 2002 for a review).  
Although some models appeal to direct
heating by the radio source (eg., Tucker \& David 1997,
Soker et al. 2000, Binney \& Kaiser 2002), the cool rims of the
cavities (McNamara et al. 2000, Fabian et al. 2000, Nulsen et al. 2002)
are inconsistent with this model (but see Brighenti \& Mathews 2002).
Some have suggested that the bubbles prevent cooling by lifting
cool, low entropy gas from the central regions of clusters
outward, where it can expand and mix with hotter
gas (David et. al. 2001, Nulsen et al.
2002, Br\"uggen \& Kaiser 2002, Churazov et al. 2002).
On the other hand, the convective lifting of material to large radii 
should erase the observed metallicity gradients in the gas
(Nulsen et al. 2002), unless the metals are rapidly replenished.

Recent theoretical studies have focussed 
on the most powerful radio sources such as Hydra A 
(Nulsen et al. 2002, Binney \& Kaiser 2002).
While $\sim 70\%$ of cooling flow cluster central
galaxies are radio audible (Burns 1990), their typical
luminosities are often one or two orders of magnitude below the
most powerful cluster sources such as Hydra A.
It has not been shown that radio sources are generally capable of
significantly reducing or quenching cooling without help from
additional heating agents (McNamara 2002).
In an interesting hybrid model, the radio source heats from the center of the
cluster while thermal conduction from the cluster's hot
halo heats the core from the outside (Ruszkowski \& Begelman 2002). 
Their hydrodynamic model appears to 
reproduce qualitatively the observed radial temperature, density, and
entropy, distributions of clusters, while producing intermittent
cooling at dramatically reduced levels.  In this section,
we test these models against the Abell 1068 data
by comparing the energy available 
from supernova explosions, the radio source, and heat conduction
to the X-ray luminosity of the cooling gas. 

The power required to reheat the
intracluster medium and maintain the gas at keV temperatures
must exceed the cooling luminosity of the gas.  
The luminosity of the gas
cooling from the ambient temperature of $\sim 4.5$ keV
through the low energy cutoff in the X-ray band
of $\sim 0.1$ keV is given by:

$$L_{\rm cool}\simeq 1.1 \times 10^{42}~\Large [{\dot M \over 1 \msunyr}\Large ]
\large [{\Delta T \over 4.4~{\rm keV} }\large ] ~{\rm erg~sec}^{-1}.$$

\noindent
For cooling at the level of  $114-145 \msunyr$ (WMM), 
we find $L_{\rm cool}\simeq 1.3-1.6\times 10^{44}~ \ergsec$.
In order to quench cooling, the total energy of all heating
mechanisms must be exceed this value, since the coupling between
the heating mechanisms and the gas will not be perfectly efficient.
Furthermore, in order to quench cooling over long timescales, the
heating must also persist over long timescales. 

\subsection{Heating by Supernova Explosions}

Assuming that star formation has proceeded
at a constant rate  of between $21\msunyr$
(the sum of the nuclear and off-nuclear constant
star formation given in Table 2) and $68 \msunyr$ (the
infrared rate), the total mass of the young population would 
be $2.1-6.7\times 10^9\msun$.
Further assuming a Type II supernova explosion rate of one
per $100\msun$ of star formation, we expect that $2.1-6.7\times 10^7$
explosions have occurred in the starburst region 
in the past 100 Myr.  Taking the total kinetic energy deposited
into the intracluster medium to be $\sim 10^{51}$ erg per explosion,
then $\sim 2.1-6.7 \times 10^{58}$ erg has been deposited by Type II 
supernovae over the past 100 Myr.  Energy input at this level would
correspond to
a feedback luminosity of $\sim 0.7-2.1\times 10^{43}~{\rm erg~sec^{-1}}$.

Drawing upon the discussion of the Type I supernova rate 
in WMM, the Type I rate integrated over 
the coolest region of the cluster is $0.03~{\rm yr}^{-1}$.  This amounts
to a feedback luminosity of $\sim 10^{42}~{\rm erg~s}^{-1}$,
less than one tenth of the energy supplied by the starburst.
The total feedback luminosity from both Type I and Type II
supernova explosions then ranges between  $\sim 8 \times 10^{42}~{\rm erg~s}^{-1}$ and 
$\sim 2.3\times 10^{43}~{\rm erg~sec^{-1}}$.  This provide up to $\sim 18\%$
of the cooling luminosity.  Therefore, supernova explosions 
can reduce the level of
cooling.  However, unless the supernova or star formation rates are 
in error by at least five times,  
they would be unable to stop cooling entirely.

\subsection{Heating by the Radio Source \& Thermal Conduction}

Now we turn to feedback from the radio source.
The monochromatic radio luminosity of the cD at
1.4 GHz is $6.26 \times 10^{39}~ \ergsec$,
based on the $FIRST$ survey image.  The radio source is barely resolved
in this image (Fig. 4), and cannot 
be separated into a core and lobe structure, if one exists
at all. The spectral slopes of radio sources can
vary widely from $\sim \nu^{-0.7}$ for radio cores to
$\sim \nu^{-1.5}$ for radio lobes.  Therefore, for simplicity
we will assume Abell 1068's radio spectrum obeys 
a power law of form $F_{\nu} \sim \nu^{-1}$ between $\nu=0.01-3$ GHz
(our conclusions are relatively insensitive to the choice of
slope). The total radio luminosity would then be
be $L_{\rm rad}=3.57\times 10^{40}~ \ergsec$.  To address the concerns
of energy deposition into the hot gas, the total mechanical
energy, not the total radiation, is relevant.
The radiative efficiency of radio jets is probably less than
unity, such that the total mechanical energy
may be a factors of $10-100$ times larger than the radio
power.  This consideration would imply a total mechanical energy of
$\sim 3.57\times 10^{41\rightarrow 42} ~\ergsec$ 
heats the intracluster medium over the lifetime
of the radio source, or up to $\sim 3\%$ of the
cooling luminosity.  

Bear in mind the crude nature 
of these calculations.  For example, the radiative efficiency of
the radio source, which we have assumed to lie between $1-10\%$, is not
well understood.  In order
to balance cooling, the radiative efficiency would have to be lowered 
to values approaching $\sim 0.02\%$, which seems implausibly low.  
Were the radiative efficiency indeed this low, the radio source
would interact strongly with the hot gas leaving
observable signatures, such as the cavities observed in other clusters.
The energy expended in $PV$ work alone would create cavities
$10-20$ kpc across.  No cavities are observed.
Furthermore, the correlation between contours of excess central
metallicity and the cD's stellar isophotes (see WMM), 
are consistent with a central metallicity gradient buildup
over the past $\sim 3$ Gyr.  This gradient suggests that
the gas has not been significantly mixed by the lifting action
of bubbles, which would act to diminish the metallicity gradient.  
Therefore, it seems unlikely that the current radio source
or others in the recent past would be capable of substantially
reducing cooling.

In Wise, McNamara, \& Murray (2003), we excluded 
heat conduction from the outer layers of the cluster as an 
important source of heat in the inner regions 
(Narayan \& Medvedev 2001, Zakamska \& Narayan 2003).
Assuming conduction is 
proceeding at the Spitzer rate, we found the inward heat
flux to be  well below the
cooling flux throughout the cluster.

In summary, heating by the radio source and thermal conduction seems
to be inconsequential. The dominant heating agent appears to be
supernova explosions in the inner 10 kpc or so associated with
the starburst.  Supernova
heating at a rate of $\sim 2\times 10^{43}~\ergsec$
should persist for $\lae$ Gyr before the 
starburst has exhausted its fuel, which would be capable 
of reducing the cooling rate by only $10\%-20\%$, provided they are
efficiently coupled to the hot gas.

\section{A Dynamically Induced Starburst?}

Although perhaps the less compelling interpretation, 
the starburst in Abell 1068 could be unrelated
to cooling, but was triggered by a tidal or ram pressure interaction.
While the optical data are too crude to distinguish
between star formation  histories of a 
dynamically-triggered starburst or a cooling flow, 
other circumstantial evidence can be cited in support 
the interaction hypothesis.   First, the star formation regions extend
well into the envelope of the cD galaxy, and galaxies {\it D} and
{\it E} appear to be associated with the off-nuclear star formation.
Galaxy {\it E} appears disrupted with blue debris
extending toward the cD.  It could originally have been a gas-rich spiral
galaxy that was tidally disrupted and ram-pressure stripped
as it traversed the high density cluster core.  Secondly, the
cD harbors a large reservoir of molecular gas (\S 3.5).  While neutral
hydrogen would have been stripped from the galaxy before it arrived 
at the center of the cluster, high density molecular
clouds tend to inhabit the tightly-bound inner regions
of galaxies.  These clouds are more likely to survive the plunge through
the cluster's halo before being ripped out by the cD and 
its associated X-ray halo as they arrive in the core.    
In addition, the Abell 1068 cD is a luminous infrared galaxy, 
and both are radio sources.  This combination of features
is commonly seen in dynamically-triggered starbursts 
(Sanders \& Mirabel 1996).  While these properties may
be circumstantial and uncompelling, they are certainly suggestive. 

Finally, inspection of Fig. 1 in WMM 
shows a filament or trail of X-ray emission extending
from the cD toward galaxy {\it E}.  The temperature
map also shows a cool, bow-shaped structure to the south-east of
the cD is reminiscent of a moving mass of cooler gas,
perhaps from a group that has just fallen into the cluster.
Filaments or trails of enhanced density and X-ray emission 
were predicted to occur during ram pressure interactions
in clusters by Stevens, Acreman, \& Ponman (1999).
The enhancements are created by gravitational
focusing in a process analogous to Bondi-Hoyle accretion,
or by the stripped debris itself.
A gravitationally-focussed trail may also be created
by the cD as it sloshes about in the
center of the cluster.  A strikingly similar situation
may be seen in Abell 1795, where an X-ray filament
$\sim 70$ kpc in length (Fabian et al. 2002) is
associated with a filament of H$\alpha$ emission (Cowie et al. 
1983) and star formation (McNamara et al. 1996).
In both instances, several disrupted satellite galaxies
are associated with the debris.  Whatever their origin,
star formation associated with the trails
contributes only a small fraction of the ultraviolet 
light from the starbursts. Therefore, the interaction
may be contributing to a centrally-condensed, cooling-fueled starburst
but is unlikely to be the dominant mechanism, unless the cold material
exchanged in the interaction quickly makes its way to
the cD's nucleus.  It is also possible that the trails
are simply cooling condensations.  Distinguishing between these 
scenarios will require
more detailed hydrodynamical simulations along the lines
of Stevens, Acreman, \& Ponman (1999) and more sensitive
ultraviolet and X-ray data.

\section{Conclusions \& Discussion}

In this paper and in the companion paper WMM, 
we presented a multiwavelength analysis 
of the Abell 1068 galaxy cluster featuring
a new {\it Chandra} X-ray image of the cluster.
We compared the levels of cold molecular gas and star formation
to the maximum cooling rates of the intracluster medium
allowed by the X-ray data.  In addition, we calculated the
energy returned to the intracluster medium from the radio
source, supernova explosions, and thermal conduction, 
and we evaluated their ability to reduce or quench cooling.
In addition, we searched for correlations 
between regions of intense star formation and
cooling condensations in the keV gas.
We used our analysis to test the self-regulated cooling flow paradigm. 
  
The near ultraviolet and infrared colors and luminosities
are consistent with either a short-duration burst of
star formation that
occurred between $6$ Myr and $40$ Myr ago,
or continuous star formation over the past 10 Myr to 100 Myr at a
rate of between $16\msunyr$ and $70\msunyr$.
The existence of a central reservoir
of $\sim 4 \times 10^{10}\msun$ of molecular gas
suggests that Abell 1068 is in the early stages
of star formation that will continue for nearly $\sim$Gyr.

The {\it Chandra} data show the sharp drop in central
gas temperature that is the characteristic signature of a cooling flow. 
The temperature drops from $\simeq 4.5$ kev beyond 
100 kpc to $\sim 2$ keV in the central starburst region of the cD.
More than 95\% of the
ultraviolet and H$\alpha$ photons from the starburst are emerging
from regions where the cooling time of the hot gas is
less than $\sim 5\times 10^8$ yr.  
The upper limit on cooling below X-ray
temperatures is $\lae 115-145\msunyr$.
This figure is several times smaller than the rate of cooling between
the ambient cluster temperature and $\sim 2$ keV, and the 
total cooling rates found with the $ROSAT$ observatory.
In this respect, Abell 1068 is similar to other clusters studied
with the {\it Chandra} and XMM observatories 
which appear to be cooling rapidly between
ambient cluster temperatures and 2 Kev, but show few of the
expected signatures of cooling to lower temperatures.
The new cooling rates agree with the star formation rates
to within factors of $2-3$, which are within the measurement 
uncertainties.  

Neither the radio source nor heat conduction are capable of
substantially reducing the level of cooling in Abell 1068.  
However, supernova explosions associated with the
starburst may be capable of reducing cooling by $\simeq 20\%$.
This result is inconsistent with new self-regulated cooling models
that are able to  maintain most of the cooling gas at keV temperatures 
through a variety of heating processes.   However, this apparent 
inconsistency may reflect the possibility that we
are observing Abell 1068 during a transient period of rapid cooling 
and intense star formation
that will eventually be slaked by a future radio outburst.

Finally, we presented circumstantial evidence
that the the starburst was triggered or
enhanced by a tidal or
ram pressure interaction with two or more neighboring galaxies. 
The X-ray and optical features in  Abell 1068 and
other clusters with dense X-ray cores are qualitatively 
similar to gravitational wakes and cool debris seen
in hydrodynamical simulations (Stevens, Acreman,
\& Ponman 1999).  As more sensitive observations in the X-ray
and the far ultraviolet bands become available,
a quantitative analysis capable of distinguishing between cooling and
stripping modes of star formation may be possible.

\acknowledgments
B.R.M would like to thank Mangala Sharma for helpful discussions.  
B.R.M. was supported by NASA Long Term Space Astrophysics Grant NAG5-11025 
and Chandra Archival Research Grant AR2-3007X.  S.S.M. was supported
by NASA grant NAS8-01130.  We thank the referee, Dr. Alastair Edge,
for carefully reading the manuscript and offering suggestions that
improved the presentation of the paper.

\clearpage

\begin{table*}
\caption{Abell 1068 Photometry}
\vspace{2.5 mm}
\baselineskip=0.01mm
\begin{tabular}{lcccccc}\hline\hline
Feature& $m(R_{\rm g})$&  $\mu_{\rm g}$&  $(U-R)_{\rm K,0}$ & $\Delta \alpha$ & $\Delta \delta$ & Area(arcsec$^2$)\\
\hline 
A   &        15.89  &           20.63  & 1.81 &0&0&78.7\\ 
B   &        17.83  &            21.67 & 2.07 &$-3.3$&6.1& 34.4\\
C   &        18.03  &           22.49 &  1.62 &$-11.8$&1.4& 60.8\\
D   &        16.94  &           21.02 &  2.32 &$-12.7$&$-7.1$&42.9\\
E  &         18.53   &          21.02 &  2.29 &$-23.0$&20.7 & 9.91\\
F  &         21.89  &           22.63 &  1.90 &$-20.2$&16.5&1.98\\
F  &         22.55  &           23.30 &  1.82 &$-16.9$&13.2&1.98\\
F  &         22.30  &           23.05 &  1.88 &$-13.6$&9.9 &1.98\\
\hline\hline
\end{tabular}
\end{table*}

\begin{table*}
\caption{Star Formation Rates \& Luminosity Masses}
\vspace{2.5 mm}
\baselineskip=0.01mm
\begin{tabular}{lccccc}\hline\hline
History& $(U-R)_{\rm AP}$& $L(U)_{\rm AP}$&  $t$& $M_{\rm AP}$ & SFR\\
\hline 
   &(mag) &$(10^{10}~{\rm L_\odot)}$ &(Myr)& ($10^8~{M}_\odot$)& ($\msunyr$)\\
\hline
          &  & Nucleus  & & &\\
\hline
Const &   $-0.3$ &      $2$ &            10&         4&               40\\
Burst &   $-0.3$ &   ... &    6&         4&              $-$\\

Const   &  $-0.1$  &       ...   &         100 &        16  &              16\\
Burst   &  $-0.1$&...    &    40&        10&              $-$ \\  
\hline 
          &  &Off Nucleus  & & &\\
\hline
Const  & $-0.2$ &0.5 & 100 & 5 & 5 \\ 
Burst & $-0.2$ &... &  7  & 0.9&$-$\\
Const  & $~~ 0.1$ &... & 500 & 15 &3 \\
Burst  & $~~0.1$ &... & 50  & 3  &$-$\\
\hline\hline
\end{tabular}
\end{table*}

\clearpage
\begin{figure}[h]
\hbox{
\hspace{.0in}
\psfig{figure=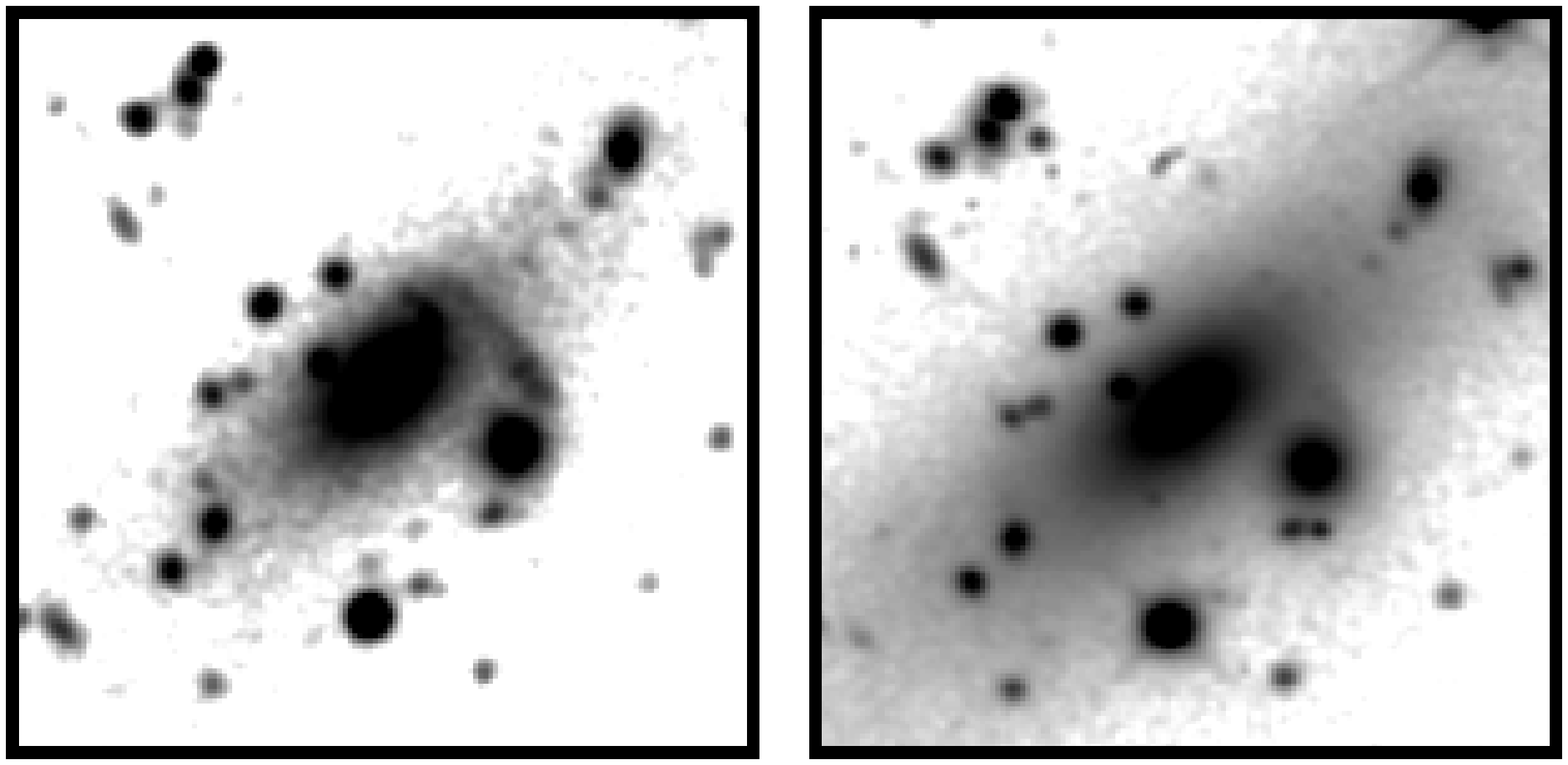,height=3.0in,width=6in,angle=270}
}
\begin{minipage}[h]{5.0truein}
Figure 1: $U$-band image (left) and $R$-band image (right) of the
central $1.2\times 1.2$ arcmin of Abell 1068. North is toward
the top; east is to the left.
\end{minipage}
\end{figure}

\clearpage
\begin{figure}[h]
\hbox{
\hspace{0.1in}
\psfig{figure=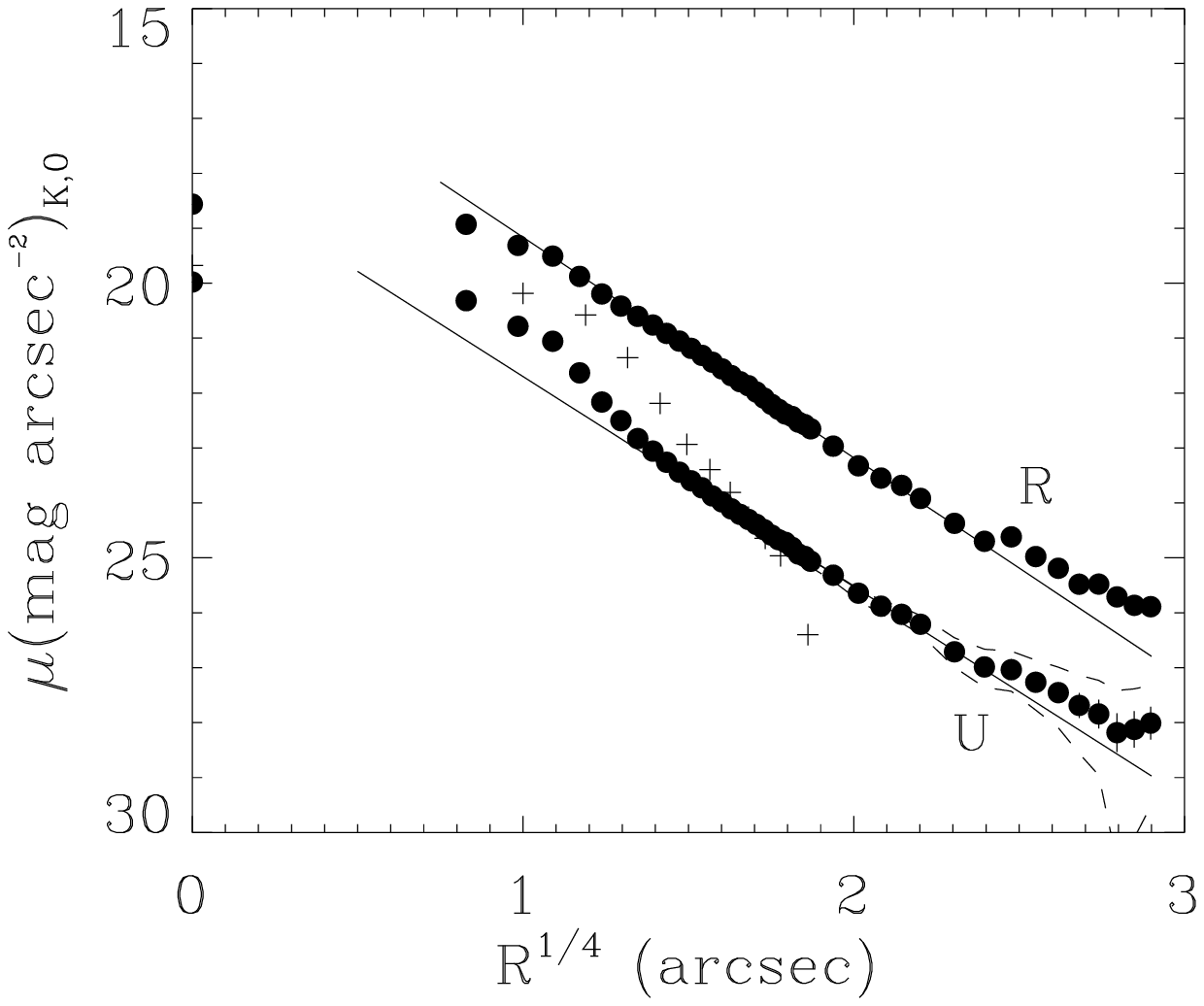,height=3.5in,width=3.5in}
\psfig{figure=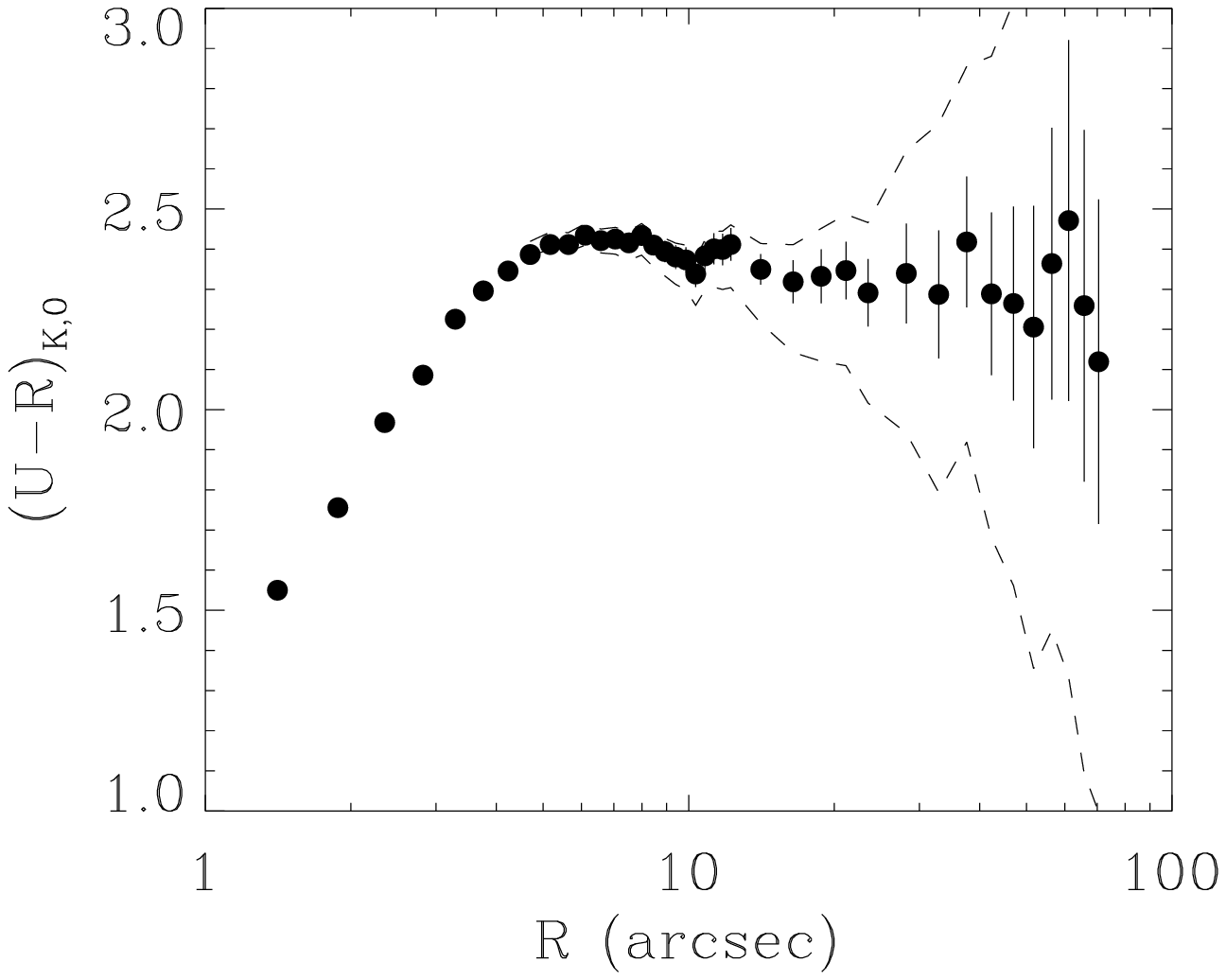,height=3.5in,width=3.5in}
}
\begin{minipage}[h]{5.0truein}
\vspace{0.1in}
Figure 2:  {\bf left} $U$ and $R$ surface brightness profiles for 
the Abell 1068 cD.  The solid lines represent $R^{1/4}$-law profiles.
The arbitrarily-scaled H$\alpha$ profile, is shown with
``+'' symbols. {\bf right} $U-R$ color profile with systematic
error envelope and statistical errors superposed. 
 
\end{minipage}
\end{figure}

\clearpage
\begin{figure}[h]
\hbox{
\hspace{0.1in}
\psfig{figure=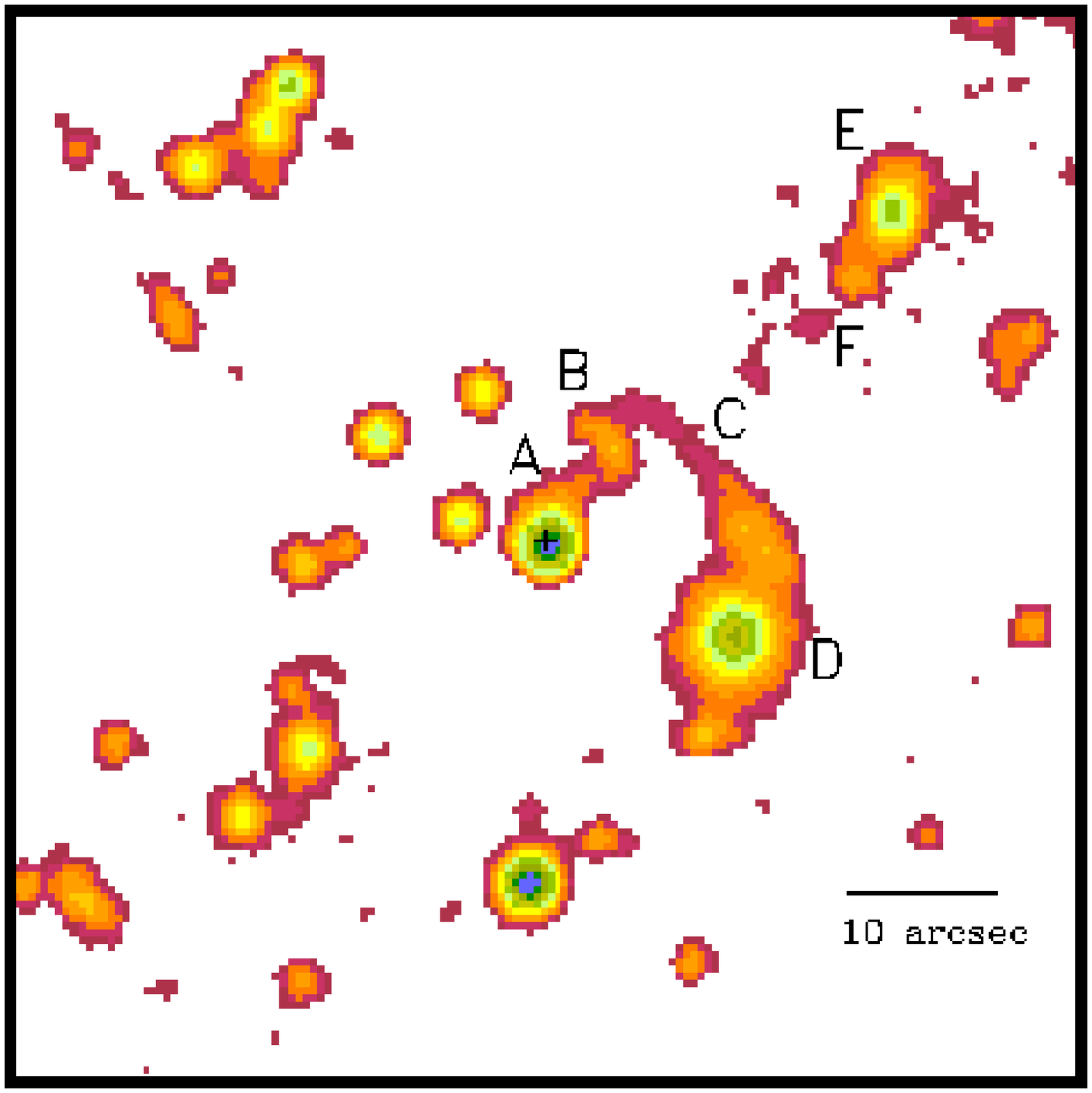,height=3.in,width=3.in,angle=90}
\psfig{figure=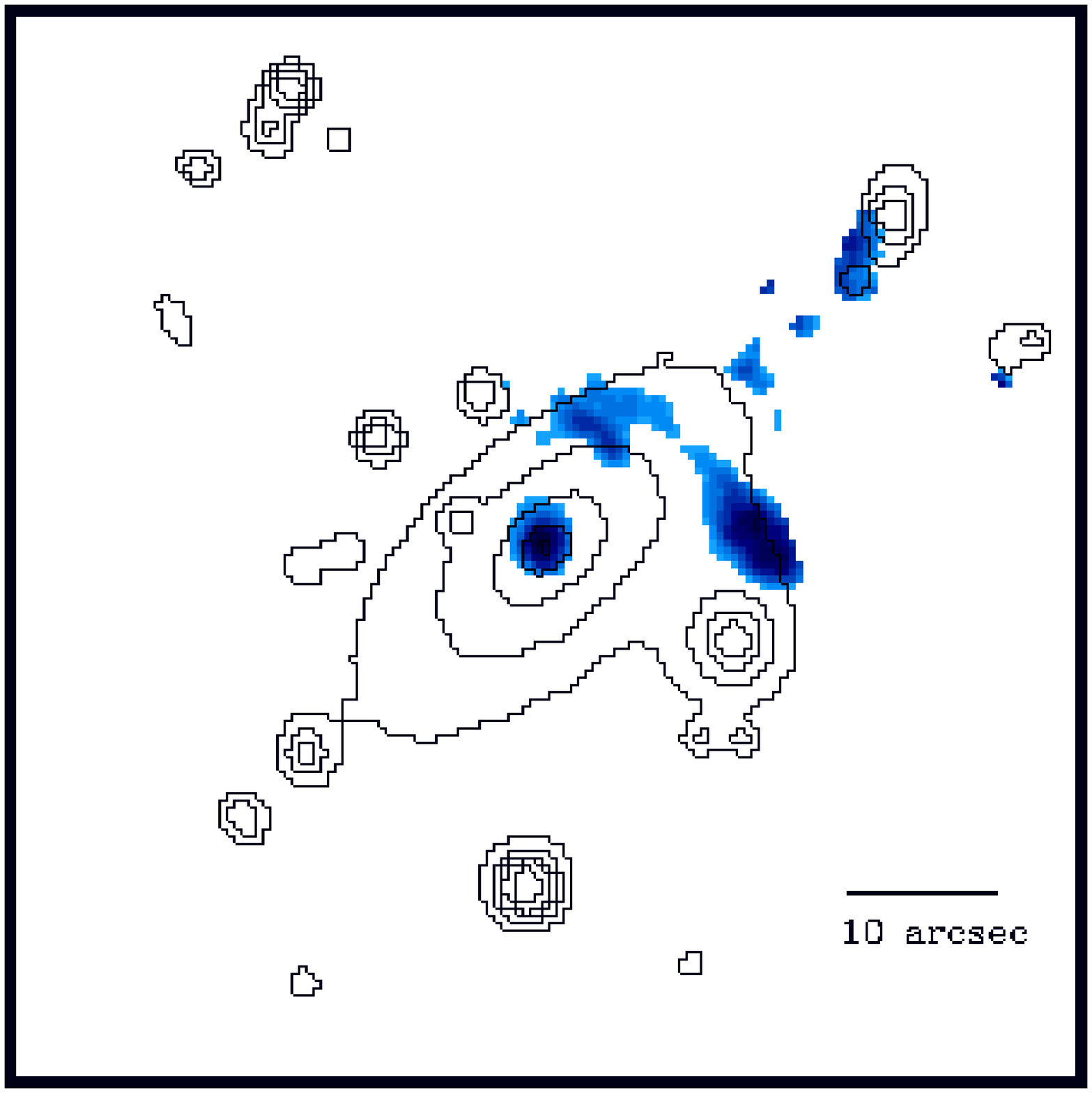,height=3.0in,width=3.0in,angle=90}
}
\begin{minipage}[h]{5.0truein}
\vspace{0.1in}
Figure 3:  {\bf left} The central $1.2\times 1.2$ arcmin
$U$-band image after subtracting the model
central galaxy. The features discussed in the text are labeled.
{\bf right} $U-R$ color map with $R$-band contours superposed. 
The anomalously blue regions are shown in blue.  North is toward
the top; east is to the left.
\end{minipage}
\end{figure}

\begin{figure}[h]
\hbox{
\hspace{0.1in}
\psfig{figure=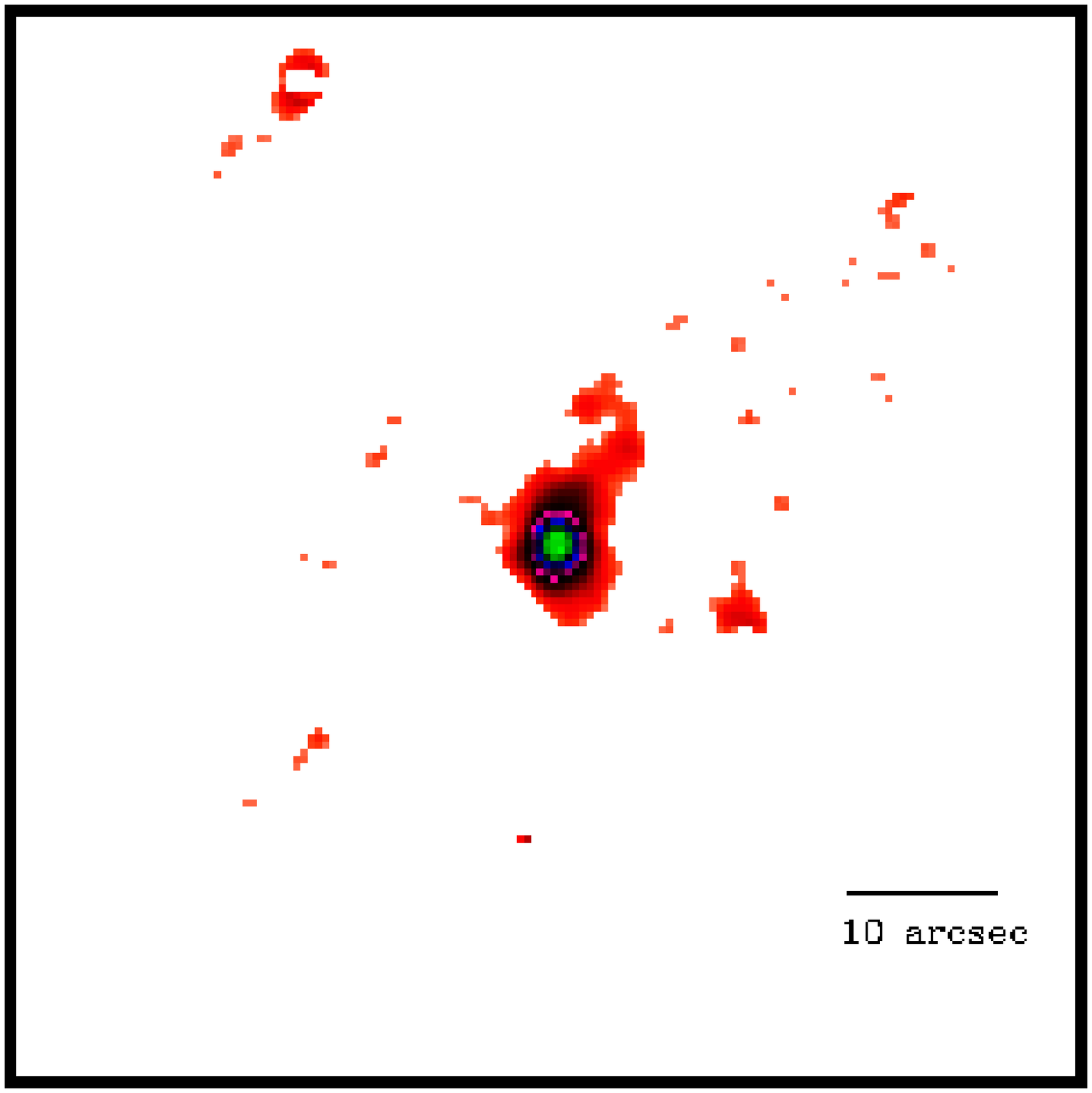,height=3.0in,width=3.in,angle=90}
\psfig{figure=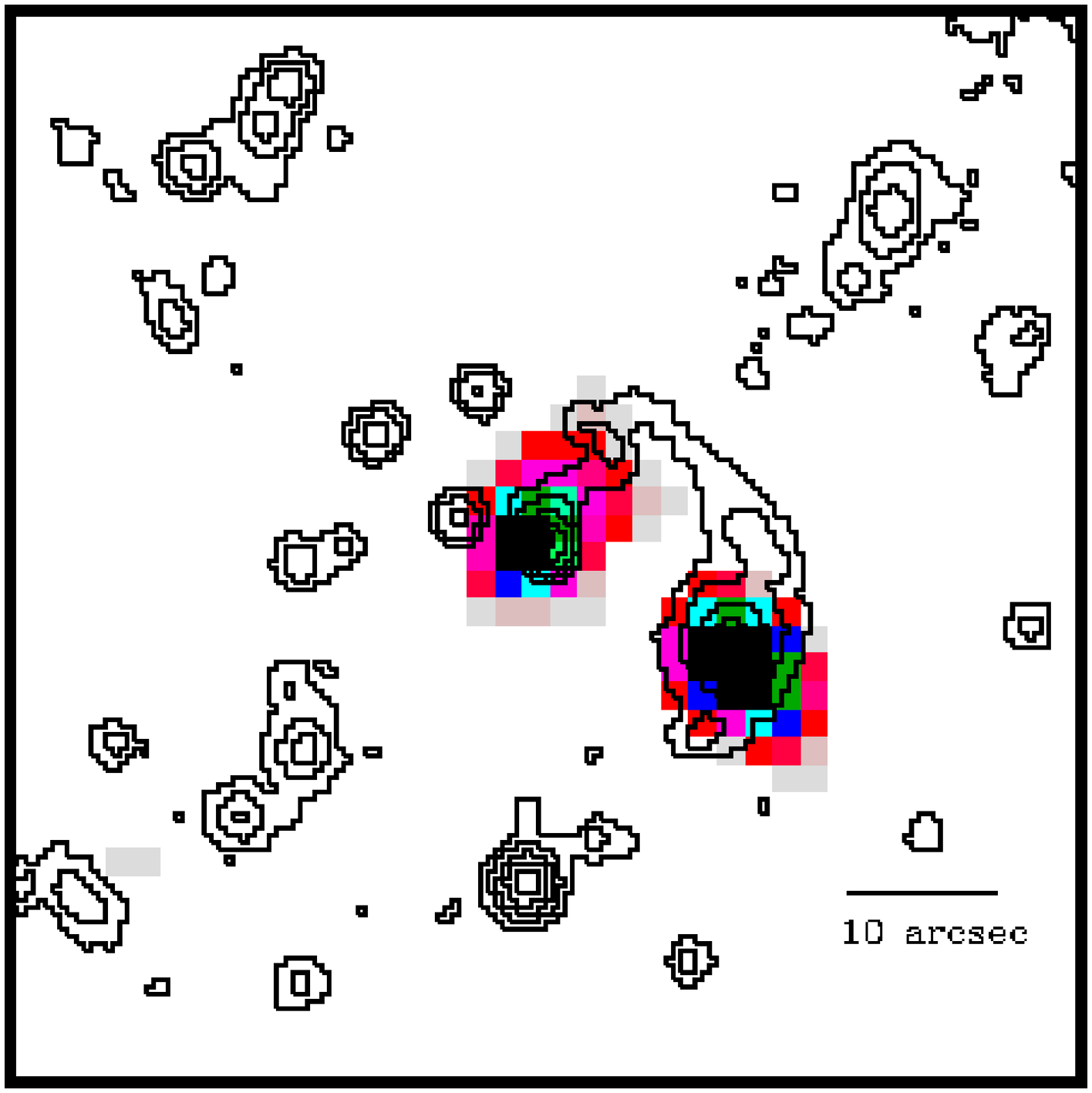,height=3.0in,width=3.in,angle=90}
}
\begin{minipage}[h]{5.0truein}
\vspace{0.1in}
Figure 4: {\bf left} The central $1.2\times 1.2$ arcmin
H$\alpha$+N [II] map shown on the same scale as
Fig. 3. {\bf right} $FIRST$ survey 20 cm radio image (color) superposed
on $U$-band contours from Fig. 3.    North is toward
the top; east is to the left.

\end{minipage}
\end{figure}

\begin{figure}[h]
\hbox{
\hspace{0.1in}
\psfig{figure=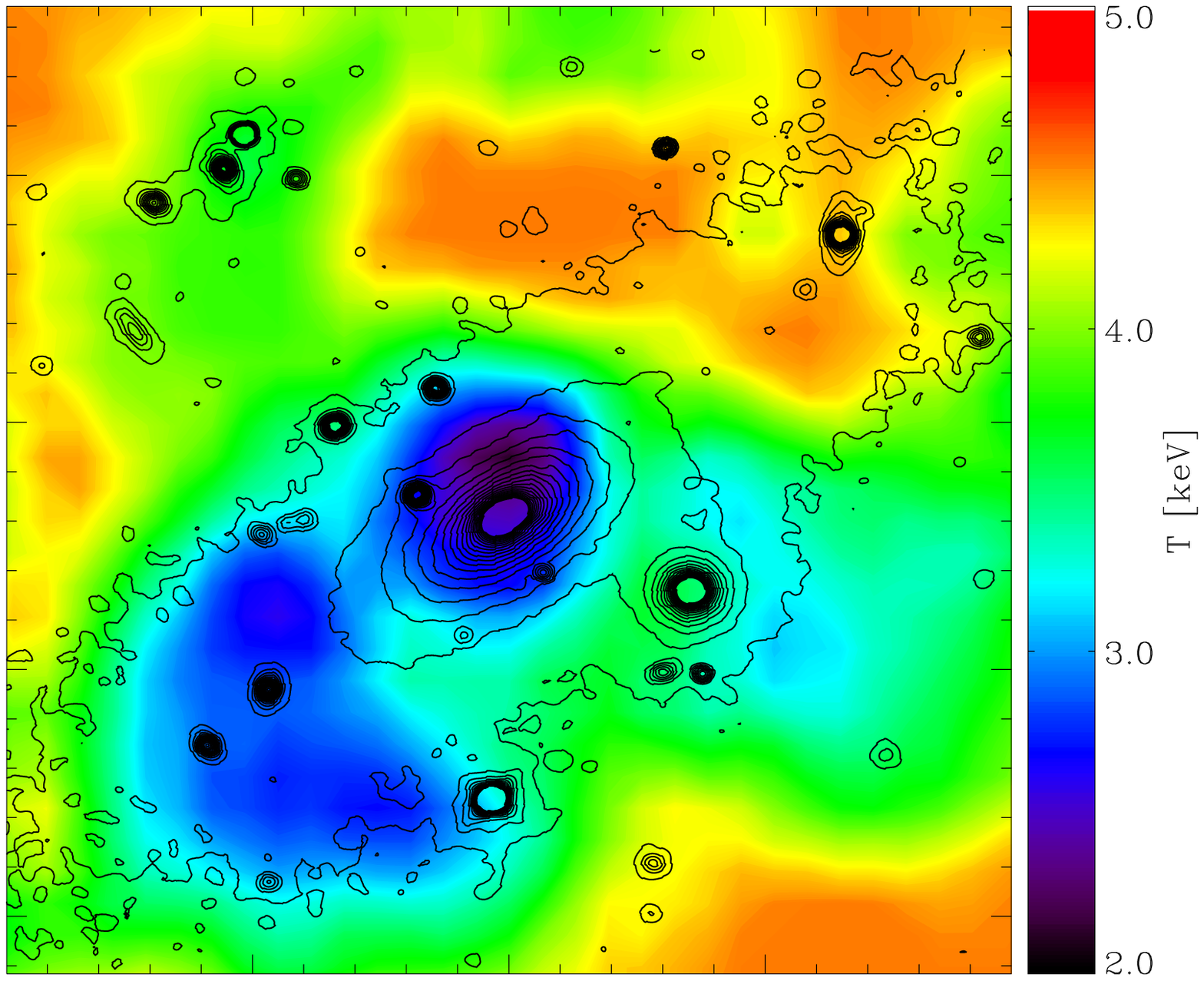,height=4.0in,width=4.5in,angle=90}
}
\begin{minipage}[h]{5.0truein}
\vspace{0.1in}
Figure 5: The central $1.2\times 1.2$ arcmin temperature 
map constructed from the Chandra data described in WMM. The
$R$-band contours showing the cD and neighboring galaxies
is superposed. North is toward the top; east is to the left. 
\end{minipage}
\end{figure}



\end{document}

%% file: brmacro.tex
\def\mathfont#1{\ifmmode{#1}\else{$#1$}\fi} 
\def\lae{\mathrel{<\kern-1.0em\lower0.9ex\hbox{$\sim$}}}  
\def\gae{\mathrel{>\kern-1.0em\lower0.9ex\hbox{$\sim$}}}  

\def\deg{\mathfont{^\circ}}  
\def\qo{\ifmmode{q_o}\else{$q_0$}\fi}  
\def\ho{\ifmmode{H_o}\else{$H_0$}\fi}   
\def\hounit{\ifmmode{{\rm km\  s}^{-1}\ {\rm Mpc}^{-        
1}}\else{${\rm    km\ s}^{-1}\ {\rm Mpc}^{-1}$}\fi}  
\def\zf{\ifmmode{z_{GF}}\else{$z_{GF}$}\fi}  
\def\kms{\ifmmode{{\rm km\ s}^{-1}}\else{${\rm km\ s}^{-1}$}\fi} 

\def\ergsec{\mathfont{ {\rm ergs\ s}^{-1}}}

\def\msun{\ifmmode{\ {\rm M}_\odot}\else{$ {\rm M}_\odot$}\fi}  
\def\msunyr{\ifmmode{\msun \ {\rm yr}^{-1}}\else{$\msun \ {\rm 
yr}^{-1}$}\fi}

\def\bv{\ifmmode{(B-V)}\else{$(B-V)$}\fi}

\def\sfr{\ifmmode{\dot S}\else{$\dot S$}\fi}

\def\ref{\noindent\hangindent=20truept\hangafter=1}

\def \vol#1          {${\bf \underline{#1}}$}